\input epsf
\input amssym 
\catcode`@=11                                   % @ is a letter here
\catcode`\|=12                                  % make sure | is not active
\catcode`\&=4                                   % and that & is alignment tab

\newcount\ncols         \ncols=\z@              % number of columns in table
\newcount\nrows         \nrows=\z@              % number of rows in table
\newcount\curcol        \curcol=\z@             % current column counter
                              % current row counter
     
\newdimen\thinsize      \thinsize=0.6pt         % thin rule width
\newdimen\thicksize     \thicksize=1.5pt        % thick rule width

\newif\iftableinfo      \tableinfotrue          % report rows and columns? Yes
\newif\ifcentertables   \centertablestrue       % center tables? Yes
%
%
%      % synonym !
     
\let\plaincr=\cr                        % save real \cr
\let\plainspan=\span                    % save real \span
\let\plaintab=&                         % save real alignment tab &
%               % to print `&' in text
\let\lparen=(                           % save left paren     
\let\NX=\noexpand                       % shorthand for \noexpand is \NX

%---------------------------------------*
% Main macros:
%       \ruledtable <table stuff> \endruledtable turns on the ruled
% table definitions and makes the table with \halign. It handles all
% the control stuff; the real work is done by \@RuledTable.
     
\def\ruledtable{\relax                          % make ruled table
    \@BeginRuledTable                           % initialize table
    \@RuledTable}%                              % now process table body

%  \@BeginRuledTable does all the work of setting things up before
% we read in the body of the table

\def\@BeginRuledTable{%                         % initialize table
   \ncols=0\nrows=0                             % reset row and column count
   \begingroup                                  % keep the following local
    \offinterlineskip                           % so vrules touch
    \def~{\phantom{0}}%                         % ~ is phantom digit
    \def\span{\plainspan\omit\relax\colcount\plainspan}%  \span USER columns
    \let\cr=\crrule                             % \cr gives a \tablerule
    \let\CR=\crthick                            % \CR gives a \thickrule
    \let\nr=\crnorule                           % \nr give no rule
    \let\|=\Vb                                  % thick vrule between columns
% 
% support old \tablestrut in place of \tstrut if it exists
%
    \ifx\tablestrut\undefined\relax             % if not defined, no problem
    \else\let\tstrut=\tablestrut\fi             % use \tablestrut
    \catcode`\|=13 \catcode`\&=13\relax         % make | and & active
    \TableActive                                % | and & get active \def`s
    \curcol=1                                   % reset column count
%
% define \Halign to do an \halign with or without a width
%
    \ifdim\tablewidth>-\maxdimen\relax          %
      \edef\@Halign{\NX\halign to \NX\tablewidth\NX\bgroup\TablePreamble}% 
      \tabskip=0pt plus 1fil                    % let things stretch
    \else                                       %
      \edef\@Halign{\NX\halign\NX\bgroup\TablePreamble}% 
      \tabskip=0pt                              % no stretch between columns
    \fi                                         %
%
% center text if needed
%
    \ifcentertables                             % should table be centered?
       \ifhmode\vskip 0pt\fi                    % yes: force vertical mode
       \line\bgroup\hss                         % center across page
    \else\hbox\bgroup                           % else: just put in \hbox
    \fi}%                                       % end of \@BeginRuledTable

%       \@RuledTable builds the table with \@Halign and getting the
% table body text as its argument.

\long\def\@RuledTable#1\endruledtable{%         % ruled table alignment
   \vrule width\thicksize                       % thick rule on side
     \vbox{\@Halign                             % then do \halign
       \thickrule                               % thick rule on top
       #1\relax                                 % body of table
       \tstrut                                  % vertical strut for last line
       \plaincr\thickrule                       % \cr, thick rule on bottom
     \egroup}%                                  % end of \halign and \vbox
   \vrule width\thicksize                       % thick rule on side, end \hbox
   \ifcentertables\hss\fi\egroup                % finish table centering
  \endgroup                                     % end group from \ruledtable
  \global\tablewidth=-\maxdimen                 %   and reset width
  \iftableinfo                                  % report rows and columns
      \immediate\write16{[Nrows=\the\nrows, Ncols=\the\ncols]}%
   \fi}%                                        % end of \@RuledTable
     
%---------------------------------------*
% Preamble and item macros:
%       This is the preamble for the \halign in \ruledtable. 
% To change how each item is processed change \TableItem.
% To make a more complicated table you can change the \TablePreamble,
% but if yo do so use the following substitutions in a standard \halign 
% preamble:
%    for  &     use     \plaintab
%    for  \cr   use     \plaincr
%    for  #     use     ##
%    for  ##    use     ####
% and put \linecount in the first column so that line counting
% works properly.

\def\TablePreamble{%                    % \ruledtable preamble
   \linecount                           % count this line
   \TableItem{####}%                    % the first item
   \plaintab\plaintab                   % && means repeat this
   \TableItem{####}%                    % the subsequent items
   \plaincr}%                           % end of preamble

%       \TableItem contains glue or spacing around the item

\def\@TableItem#1{%                             % centers item in ruled table
   \hfil\tablespace                             % left glue
   #1\relax                                     % item
   \tablespace\hfil                             % right glue
    }%

\def\@tableright#1{%                    % right justifies item in ruled table
   \hfil\tablespace\relax               % left glue
   #1\relax                             % item
   \tablespace\relax}%                  % right glue

\def\@tableleft#1{%                     % left justifies item in ruled table
   \tablespace\relax                    % left glue
   #1\relax                             % item
   \tablespace\hfil}%                   % right glue

\let\TableItem=\@TableItem              % default is centered
     
\def\RightJustifyTables{\let\TableItem=\@tableright}%   % to right justify
\def\LeftJustifyTables{\let\TableItem=\@tableleft}%     % to left justify
\def\NoJustifyTables{\let\TableItem=\@TableItem}%       % to center

\def\LooseTables{\let\tablespace=\quad}%        % table spacing is \quad
\def\TightTables{\let\tablespace=\space}%       % table spacing is space
\LooseTables                                    % default is \quad

%---------------------------------------*
% Table Height and Width:
%
%  Normally tables are set to their natural width.  If \tablewidth
%  has been set then we set the table to that width instead,
%  but only for the next table.  Then \tablewidth is turned off

\newdimen\tablewidth    \tablewidth=-\maxdimen  % start ``turned off''

%      \setRuledStrut sets up the vertical strut \tstrut with
% the appropriate dimensions to hold up one line of a ruled table. 

\def\setRuledStrut{% sets interlines spacing for ruled tables
   \dimen@=\baselineskip                        % \dimen@ = extra space
   \advance\dimen@ by-\normalbaselineskip       % between lines
   \ifdim\dimen@<.5ex \dimen@=.5ex\fi           % minimum space
   \setbox0=\hbox{\lparen}%                     % get character size
   \dimen1=\dimen@ \advance\dimen1 by \ht0      % space above line
   \dimen2=\dimen@ \advance\dimen2 by \dp0      % space below line
   \def\tstrut{\vrule height\dimen1 depth\dimen2 width\z@}%
   }%

\def\tstrut{\vrule height 3.1ex depth 1.2ex width 0pt}%  default

%      \tstrut does not produce the correct spacing if the entry in
% the table is too high. The following constructs a strut higher than its
% argument and then prints the argument. The minimum space value should
% be the same as in \setRuledStrut. 

\def\bigitem#1{%                                % larger table entry
   \setbox0=\hbox{#1}%                          % put arg. in box and
   \dimen1 =\ht0 \dimen2 =\dp0                  % get its size
   \dimen@ =\baselines@ve                       % \dimen@ = extra space
   \advance\dimen@ by-\normalbaselineskip       %   between lines
   \ifdim\dimen@<.25ex \dimen@=.25ex\fi         % minimum space
   \advance\dimen1 by \dimen@                   % space above line
   \advance\dimen2 by \dimen@                   % space below line
   \vrule height\dimen1 depth\dimen2 width\z@   % make strut to size
   \copy0}%                                     % print argument

%       \vctr{stuff} centers the stuff vertically, so that it can
% appear between two ROWS.
     
%

%---------------------------------------*
% Vertical rules:
%       \tab, \vb and \Vb are used in a table to separate columns with no
% rule, a thin rule, or a thick rule, respectively.  \nextcolumn{<rule>} skips
% to  the next column and puts the <rule> between columns.  Use this to build
% your own separators.
     
\def\nextcolumn#1{%                             % move to next col.
   \plaintab\omit#1\relax\colcount              % tab, insert #1, count
   \plaintab}%                                  % tab to next user col.
     
\def\tab{%                                      % no rule between columns
   \nextcolumn{\relax}}%                        % count column

                                  % synonym for \tab

\def\vb{%                                       % thin rule between columns
   \nextcolumn{\vrule width\thinsize}}%         % count and rule

\def\Vb{%                                       % THICK rule between columns
   \nextcolumn{\vrule width\thicksize}}%        % count and thick rule

\def\dbl{%                                      % double rule between columns
   \nextcolumn{\vrule width\thinsize            % count and rule
   \hskip\thinsize\vrule width\thinsize}}%      % and skip and rule

%       \TableActive makes | the same as \vb and & the same as \tab so
% these single characters can be used between columns. Here we have to
% make & and | active so we get the active version of the characters,
% hence the temporary change of \catcode.
     
{\catcode`\|=13 \let|0
 \catcode`\&=13 \let&0
 \gdef\TableActive{\let|=\vb \let&=\tab}%
}% end \catcode`s

%---------------------------------------*
% Horizontal rules:
%       These replacements for \cr put a wide \vrule at the end of the
% line and maybe put a rule under the line, then begin the next line
% with the wide \vrule from the preamble.

\def\crrule{\relax                      % \cr plus rule
   \tstrut                              % strut for spacing
   \plaincr\tablerule                   % \cr, regular rule below line
  }%

\def\crthick{\relax                     % \cr plus thick rule
   \tstrut                              % strut for vertical spacing
   \plaincr\thickrule                   % \cr, rule, begin next line
  }%                
     
\def\crnorule{\relax                    % \cr plus no rule
   \tstrut                              % strut for spacing
   \plaincr                             % \cr, norule, begin next line
   }%
   
%                  % for partial rules, no strut

%       These rules go across the table.
     
\def\tablerule{\noalign{\hrule height\thinsize depth 0pt}}%
\def\thickrule{\noalign{\hrule height\thicksize depth 0pt}}%

%       Rules for individual columns. You must use \cskip in columns
% with no rules to \omit the \TablePreamble.

%
%
%
     
%---------------------------------------*
% Counting macros:     
%       These macros count rows and columns of the table. After the
% first line has been processed \the\ncols is the total number of
% columns in the table, which may be useful. During processing,
% \the\curcol is the number of the current column, while \the\currow is
% the number of the current row.

\def\linecount{\relax\global\ncols=\curcol      % save column count in \ncols
   \global\curcol=1                             % and reset counter
   \global\advance\nrows by 1\relax}%           % and advance row counter
     
\def\colcount{\relax                            %
   \global\advance\curcol by 1\relax}%          % advance column counter

%---------------------------------------*
% TEXT TABLES.
%  To put text in a table use \para{<text>}, with \parasize set
%  to the desired width of the text.

\newdimen\parasize      \parasize=4in           % paragraph size in tables

%

%---------------------------------------*
% TABLES.TEX
%   For compatability with Cowan's TABLES.TEX we also allow the
% syntax \begintable ... \endtable, which do the same thing.
%

\def\begintable{\relax                          % make ruled table
    \@BeginRuledTable                           % initialize table
    \@begintable}%                              % now process table body

\long\def\@begintable#1\endtable{%              % ruled table alignment
   \@RuledTable#1\endruledtable}%               % same as \ruledtable

%---------------------------------------*
% Turn off @ as letter before we exit

\catcode`@=12                                   % @ is not a letter

%>>> EOF TXSruled.tex <<<

%\input draft

%%%%%%%%%%%%%%%%%%%%%%%%%%%%%%%%%%%%%%%%%%%%%%%%%%%%%%%%%%%%%%%%%
%                                                               %
%       FONT FAMILIES:                                          %
%                                                               %
%%%%%%%%%%%%%%%%%%%%%%%%%%%%%%%%%%%%%%%%%%%%%%%%%%%%%%%%%%%%%%%%%
%                                                               %
%       Define script letters as rsfs                           %
%               (or redefine as cal)                            %
%                                                               %
%                                                               %
%%%%%%%%%%%%%%%%%%%%%%%%%%%%%%%%%%%%%%%%%%%%%%%%%%%%%%%%%%%%%%%%%
\newfam\scrfam
\batchmode\font\tenscr=rsfs10 \errorstopmode
\ifx\tenscr\nullfont
        \message{rsfs script font not available. Replacing with calligraphic.}
        \def\scr{\cal}
\else   
        \font\sevenscr=rsfs7
        \font\fivescr=rsfs5
        \skewchar\tenscr='177 \skewchar\sevenscr='177 \skewchar\fivescr='177
        \textfont\scrfam=\tenscr \scriptfont\scrfam=\sevenscr
        \scriptscriptfont\scrfam=\fivescr
        \def\scr{\fam\scrfam}
        \def\cal{\scr}
\fi
%%%%%%%%%%%%%%%%%%%%%%%%%%%%%%%%%%%%%%%%%%%%%%%%%%%%%%%%%%%%%%%%%
%                                                               %
%       fraktur (or redefine as italic)                     %
%                                                               %
%%%%%%%%%%%%%%%%%%%%%%%%%%%%%%%%%%%%%%%%%%%%%%%%%%%%%%%%%%%%%%%%%
\catcode`\@=11
\newfam\frakfam
\batchmode\font\tenfrak=eufm10 \errorstopmode
\ifx\tenfrak\nullfont
        \message{eufm font not available. Replacing with italic.}
        
\else
    
    \font\sevenfrak=eufm7 \font\fivefrak=eufm5
    \textfont\frakfam=\tenfrak
    \scriptfont\frakfam=\sevenfrak \scriptscriptfont\frakfam=\fivefrak
    
\fi
\catcode`\@=\active
%%%%%%%%%%%%%%%%%%%%%%%%%%%%%%%%%%%%%%%%%%%%%%%%%%%%%%%%%%%%%%%%%
%                                                               %
%       Blackboard bold (or redefine as boldface)               %
%                                                               %
%%%%%%%%%%%%%%%%%%%%%%%%%%%%%%%%%%%%%%%%%%%%%%%%%%%%%%%%%%%%%%%%%
\newfam\msbfam
\batchmode\font\twelvemsb=msbm10 scaled\magstep1 \errorstopmode
\ifx\twelvemsb\nullfont\def\Bbb{\bf}

    \message{Blackboard bold not available. Replacing with boldface.}
\else   \catcode`\@=11
        \font\tenmsb=msbm10 \font\sevenmsb=msbm7 \font\fivemsb=msbm5
        \textfont\msbfam=\tenmsb
        \scriptfont\msbfam=\sevenmsb \scriptscriptfont\msbfam=\fivemsb
        \def\Bbb{\relax\expandafter\Bbb@}
        \def\Bbb@#1{{\Bbb@@{#1}}}
        \def\Bbb@@#1{\fam\msbfam\relax#1}
        \catcode`\@=\active

\fi
%%%%%%%%%%%%%%%%%%%%%%%%%%%%%%%%%%%%%%%%%%%%%%%%%%%%%%%%%%%%%%%%%
%                                                               %
%       Capitals                                                %
%                                                               %
%%%%%%%%%%%%%%%%%%%%%%%%%%%%%%%%%%%%%%%%%%%%%%%%%%%%%%%%%%%%%%%%%
\newfam\cpfam
\def\sectionfonts{\relax
    \textfont0=\twelvecp          \scriptfont0=\ninecp
      \scriptscriptfont0=\sixrm
    \textfont1=\twelvei           \scriptfont1=\ninei
      \scriptscriptfont1=\sixi
    \textfont2=\twelvesy           \scriptfont2=\ninesy
      \scriptscriptfont2=\sixsy
    \textfont3=\twelveex          \scriptfont3=\tenex
      \scriptscriptfont3=\tenex
    \textfont\itfam=\twelveit     \scriptfont\itfam=\nineit
    \textfont\slfam=\twelvesl     \scriptfont\slfam=\ninesl
    \textfont\bffam=\twelvebf     \scriptfont\bffam=\ninebf
      \scriptscriptfont\bffam=\sixbf
    \textfont\ttfam=\twelvett
    \textfont\cpfam=\twelvecp
}
%%%%%%%%%%%%%%%%%%%%%%%%%%%%%%%%%%%%%%%%%%%%%%%%%%%%%%%%%%%%%%%%%
%                                                               %
%       More FONTS:                                             %
%                                                               %
%%%%%%%%%%%%%%%%%%%%%%%%%%%%%%%%%%%%%%%%%%%%%%%%%%%%%%%%%%%%%%%%%
        \font\eightrm=cmr8              \def\xrm{\eightrm}
        \font\eightbf=cmbx8             \def\xbf{\eightbf}
        \font\eightit=cmti10 at 8pt     \def\xit{\eightit}
                       
        \font\sixrm=cmr6                
%%%     \font\eightit=cmti8             \def\xit{\eightit}
        \font\eighttt=cmtt8             
        \font\eightcp=cmcsc8
        \font\eighti=cmmi8              \def\xold{\eighti}
        \font\eightib=cmmib8             \def\xbold{\eightib}
        \font\teni=cmmi10               \def\old{\teni}
        \font\ninei=cmmi9
        \font\tencp=cmcsc10
        \font\ninecp=cmcsc9

        \font\twelvei=cmmi12
        \font\twelvecp=cmcsc10 scaled\magstep1
        
        \font\fiverm=cmr5
        
        \font\twelvesy=cmsy12
        \font\ninesy=cmsy9
        \font\sixsy=cmsy6
        \font\twelveex=cmex12

        \font\twelveit=cmti12
        \font\nineit=cmti9
        
        \font\twelvesl=cmsl12
        \font\ninesl=cmsl9
        
        \font\twelvebf=cmbx12
        \font\ninebf=cmbx9
        \font\sixbf=cmbx6
        \font\twelvett=cmtt12

        \font\sixi=cmmi6

\batchmode\font\tenhelvbold=phvb at10pt \errorstopmode
\ifx\tenhelvbold\nullfont
        \message{phvb font not available. Replacing with cmr.}
    \font\tenhelvbold=cmb10   
    \font\twelvehelvbold=cmb12
    
    \font\sixteenhelvbold=cmb16
  \else
    \font\tenhelvbold=phvb at10pt   
    \font\twelvehelvbold=phvb at12pt
     at14pt
    \font\sixteenhelvbold=phvb at16pt
\fi

\def\noblackbox{\overfullrule=0pt}
\noblackbox

\font\eightmi=cmmi8
\font\sixmi=cmmi6
\font\fivemi=cmmi5

\font\eightsy=cmsy8
\font\sixsy=cmsy6
\font\fivesy=cmsy5

\font\eightsl=cmsl8

%=========================================================================
% {\eightpoint\rm ...}  { all fonts will be 8 point including math fonts }
% {\eightpoint $...$ }  { math only to be in 8 point font }
%-------------------------------------------------------------------------
\def\eightpoint{
\def\rm{\fam0\eightrm}
\textfont0=\eightrm \scriptfont0=\sixrm \scriptscriptfont0=\fiverm
\textfont1=\eightmi  \scriptfont1=\sixmi  \scriptscriptfont1=\fivemi
\textfont2=\eightsy \scriptfont2=\sixsy \scriptscriptfont2=\fivesy
\textfont3=\tenex   \scriptfont3=\tenex \scriptscriptfont3=\tenex
\textfont\itfam=\eightit \def\it{\fam\itfam\eightit}
\textfont\slfam=\eightsl \def\sl{\fam\slfam\eightsl}
\textfont\ttfam=\eighttt \def\tt{\fam\ttfam\eighttt}
\textfont\bffam=\eightbf \scriptfont\bffam=\sixbf 
                         \scriptscriptfont\bffam=\fivebf
                         \def\bf{\fam\bffam\eightbf}
\normalbaselineskip=10pt}

%%%%%%%%%%%%%%%%%%%%%%%%%%%%%%%%%%%%%%%%%%%%%%%%%%%%%%%%%%%%%%%%%
%                                                               %
%       HEADLINE:                                               %
%                                                               %
%%%%%%%%%%%%%%%%%%%%%%%%%%%%%%%%%%%%%%%%%%%%%%%%%%%%%%%%%%%%%%%%%
\newtoks\headtext
\headline={\ifnum\pageno=1\hfill\else
    \ifodd\pageno{\eightcp\the\headtext}{ }\dotfill{ }{\old\folio}
    \else{\old\folio}{ }\dotfill{ }{\eightcp\the\headtext}\fi
    \fi}
\def\makeheadline{\vbox to 0pt{\vss\noindent\the\headline\break
\hbox to\hsize{\hfill}}
        \vskip2\baselineskip}
%%%%%%%%%%%%%%%%%%%%%%%%%%%%%%%%%%%%%%%%%%%%%%%%%%%%%%%%%%%%%%%%%
%                                                               %
%       FOOTNOTES:                                              %
%                                                               %
%%%%%%%%%%%%%%%%%%%%%%%%%%%%%%%%%%%%%%%%%%%%%%%%%%%%%%%%%%%%%%%%%
\newcount\infootnote
\infootnote=0
\def\foot#1#2{\infootnote=1
\footnote{${}^{#1}$}{\vtop{\baselineskip=.75\baselineskip
\advance\hsize by
-\parindent{\eightpoint\rm\hskip-\parindent #2}\hfill\vskip\parskip}}\infootnote=0$\,$}
%%%%%%%%%%%%%%%%%%%%%%%%%%%%%%%%%%%%%%%%%%%%%%%%%%%%%%%%%%%%%%%%%
%                                                               %
%       REFERENCES:                                             %
%                                                               %
%%%%%%%%%%%%%%%%%%%%%%%%%%%%%%%%%%%%%%%%%%%%%%%%%%%%%%%%%%%%%%%%%
\newcount\refcount
\refcount=1
\newwrite\refwrite
\def\oldsize{\ifnum\infootnote=1\xold\else\old\fi}
\def\ref#1#2{
    \def#1{{{\oldsize\the\refcount}}\ifnum\the\refcount=1\immediate\openout\refwrite=\jobname.refs\fi\immediate\write\refwrite{\item{[{\xold\the\refcount}]}
    #2\hfill\par\vskip-2pt}\xdef#1{{\noexpand\oldsize\the\refcount}}\global\advance\refcount by 1}
    }
\def\refout{\catcode`\@=11
        \xrm\immediate\closeout\refwrite
        \vskip2\baselineskip
        {\noindent\twelvecp References}\hfill
        \par\nobreak\vskip\baselineskip
        %\vskip.25\baselineskip%%%%
        %\parskip=.875\parskip
        %\baselineskip=.8\baselineskip
        \baselineskip=.75\baselineskip
        \input\jobname.refs
        %\parskip=8\parskip \divide\parskip by 7
        %\baselineskip=1.25\baselineskip
        \baselineskip=4\baselineskip \divide\baselineskip by 3
        \catcode`\@=\active\rm}

\def\hepth#1{\href{http://arxiv.org/abs/hep-th/#1}{arXiv:hep-th/{\xold#1}}}

\def\arxiv#1#2{\href{http://arxiv.org/abs/#1.#2}{arXiv:{\xold#1}.{\xold#2}}}
\def\jhep#1#2#3#4{\href{http://jhep.sissa.it/stdsearch?paper=#2\%28#3\%29#4}{J. High Energy Phys. {\xbold #1#2} ({\xold#3}) {\xold#4}}}

\def\CQG#1#2#3{Class. Quantum Grav. {\xbold#1} ({\xold#2}) {\xold#3}}

\def\JHEP{\jhep}

\def\LMP#1#2#3{Lett. Math. Phys. {\xbold#1} ({\xold#2}) {\xold#3}}

\def\NPB#1#2#3{Nucl. Phys. {\xbf B}{\xbold#1} ({\xold#2}) {\xold#3}}

\def\PLB#1#2#3{Phys. Lett. {\xbf B}{\xbold#1} ({\xold#2}) {\xold#3}}

\def\PRD#1#2#3{Phys. Rev. {\xbf D}{\xbold#1} ({\xold#2}) {\xold#3}}
\def\PRL#1#2#3{Phys. Rev. Lett. {\xbold#1} ({\xold#2}) {\xold#3}}

%%%%%%%%%%%%%%%%%%%%%%%%%%%%%%%%%%%%%%%%%%%%%%%%%%%%%%%%%%%%%%%%%
%                                                               %
%       SECTION NUMBERING:                                      %
%                                                               %
%%%%%%%%%%%%%%%%%%%%%%%%%%%%%%%%%%%%%%%%%%%%%%%%%%%%%%%%%%%%%%%%%
\newcount\sectioncount
\sectioncount=0
\def\section#1#2{\global\eqcount=0
    \global\subsectioncount=0
        \global\advance\sectioncount by 1
    \ifnum\sectioncount>1
            \vskip2\baselineskip
    \fi
    \noindent
%        \line{\twelvecp\the\sectioncount. #2\hfill}
       \line{\sectionfonts\twelvecp\the\sectioncount. #2\hfill}
        \par\nobreak\vskip.8\baselineskip\noindent
        \xdef#1{{\old\the\sectioncount}}}
\newcount\subsectioncount
\def\subsection#1#2{\global\advance\subsectioncount by 1
    \par\nobreak\vskip.8\baselineskip\noindent
    \line{\tencp\the\sectioncount.\the\subsectioncount. #2\hfill}
    \vskip.5\baselineskip\noindent
    \xdef#1{{\old\the\sectioncount}.{\old\the\subsectioncount}}}
\newcount\appendixcount
\appendixcount=0
\def\appendix#1{\global\eqcount=0
        \global\advance\appendixcount by 1
        \vskip2\baselineskip\noindent
        \ifnum\the\appendixcount=1
        \hbox{\twelvecp Appendix A: #1\hfill}
        \par\nobreak\vskip\baselineskip\noindent\fi
    \ifnum\the\appendixcount=2
        \hbox{\twelvecp Appendix B: #1\hfill}
        \par\nobreak\vskip\baselineskip\noindent\fi
    \ifnum\the\appendixcount=3
        \hbox{\twelvecp Appendix C: #1\hfill}
        \par\nobreak\vskip\baselineskip\noindent\fi}
\def\acknowledgements{\vskip2\baselineskip\noindent
        \underbar{\it Acknowledgements:}\ }
%%%%%%%%%%%%%%%%%%%%%%%%%%%%%%%%%%%%%%%%%%%%%%%%%%%%%%%%%%%%%%%%%
%                                                               %
%       EQUATION NUMBERING                                      %
%                                                               %
%%%%%%%%%%%%%%%%%%%%%%%%%%%%%%%%%%%%%%%%%%%%%%%%%%%%%%%%%%%%%%%%%
\newcount\eqcount
\eqcount=0
\def\Eqn#1{\global\advance\eqcount by 1
\ifnum\the\sectioncount=0
    \xdef#1{{\old\the\eqcount}}
    \eqno({\oldstyle\the\eqcount})
\else
        \ifnum\the\appendixcount=0
            \xdef#1{{\old\the\sectioncount}.{\old\the\eqcount}}
                \eqno({\oldstyle\the\sectioncount}.{\oldstyle\the\eqcount})\fi
        \ifnum\the\appendixcount=1
            \xdef#1{{\oldstyle A}.{\old\the\eqcount}}
                \eqno({\oldstyle A}.{\oldstyle\the\eqcount})\fi
        \ifnum\the\appendixcount=2
            \xdef#1{{\oldstyle B}.{\old\the\eqcount}}
                \eqno({\oldstyle B}.{\oldstyle\the\eqcount})\fi
        \ifnum\the\appendixcount=3
            \xdef#1{{\oldstyle C}.{\old\the\eqcount}}
                \eqno({\oldstyle C}.{\oldstyle\the\eqcount})\fi
\fi}
\def\eqn{\global\advance\eqcount by 1
\ifnum\the\sectioncount=0
    \eqno({\oldstyle\the\eqcount})
\else
        \ifnum\the\appendixcount=0
                \eqno({\oldstyle\the\sectioncount}.{\oldstyle\the\eqcount})\fi
        \ifnum\the\appendixcount=1
                \eqno({\oldstyle A}.{\oldstyle\the\eqcount})\fi
        \ifnum\the\appendixcount=2
                \eqno({\oldstyle B}.{\oldstyle\the\eqcount})\fi
        \ifnum\the\appendixcount=3
                \eqno({\oldstyle C}.{\oldstyle\the\eqcount})\fi
\fi}
\def\multi{\global\advance\eqcount by 1}
\def\multieq#1#2{
    \ifnum\the\sectioncount=0
        \eqno({\oldstyle\the\eqcount})
         \xdef#1{{\old\the\eqcount#2}}
    \else
        \xdef#1{{\old\the\sectioncount}.{\old\the\eqcount}#2}
        \eqno{({\oldstyle\the\sectioncount}.{\oldstyle\the\eqcount}#2)}
    \fi}

%%%%%%%%%%%%%%%%%%%%%%%%%%%%%%%%%%%%%%%%%%%%%%%%%%%%%%%%%%%%%%%%%
%                                                               %
%       Hyperrefs:                                          %
%                                                               %
%%%%%%%%%%%%%%%%%%%%%%%%%%%%%%%%%%%%%%%%%%%%%%%%%%%%%%%%%%%%%%%%%
\newtoks\url
\def\Href#1#2{\catcode`\#=12\url={#1}\catcode`\#=\active#2}
\def\href#1#2{{#2}}
\def\hhref#1{{#1}}
%%%%%%%%%%%%%%%%%%%%%%%%%%%%%%%%%%%%%%%%%%%%%%%%%%%%%%%%%%%%%%%%%
%                                                               %
%       FORMAT:                                                 %
%                                                               %
%%%%%%%%%%%%%%%%%%%%%%%%%%%%%%%%%%%%%%%%%%%%%%%%%%%%%%%%%%%%%%%%%
\parskip=3.5pt plus .3pt minus .3pt
\baselineskip=14pt plus .1pt minus .05pt
\lineskip=.5pt plus .05pt minus .05pt
\lineskiplimit=.5pt
\abovedisplayskip=18pt plus 4pt minus 2pt
\belowdisplayskip=\abovedisplayskip
\hsize=14cm
\vsize=20.8cm
\hoffset=1.5cm
\voffset=1.5cm
\frenchspacing
\footline={}
\raggedbottom
%%%%%%%%%%%%%%%%%%%%%%%%%%%%%%%%%%%%%%%%%%%%%%%%%%%%%%%%%%%%%%%%%
%                                                               %
%       VARIOUS DEFINITIONS                                     %
%                                                               %
%%%%%%%%%%%%%%%%%%%%%%%%%%%%%%%%%%%%%%%%%%%%%%%%%%%%%%%%%%%%%%%%%

\def\ss{\scriptstyle}
\def\sss{\scriptscriptstyle}
\def\*{\partial}
\def\punkt{\,\,.}
\def\komma{\,\,,}

\def\={\!=\!}
\def\small#1{{\hbox{$#1$}}}

\def\fraction#1{\small{1\over#1}}
\def\fr{\fraction}
\def\Fraction#1#2{\small{#1\over#2}}
\def\Fr{\Fraction}

\def\eg{{\tenit e.g.}}

\def\ie{{\tenit i.e.}}

\def\a{\alpha}
\def\b{\beta}

\def\d{\delta}
\def\e{\varepsilon}
\def\g{\gamma}

\def\m{\mu}

\def\ra{\rightarrow}

%%%%%%%%%%%%%%%%%%%%%%%%%%%%%%%%%%%%%%%%%%%%%%%%%%%%%%%%%%%%%%%%%%%%%%%%%%%%%

%\def\rdiag#1{\hbox{\epsfxsize=12pt\lower3pt\hbox{\epsffile{#1.eps}}}}
%\def\rrrdiag#1{\hbox{\epsfxsize=12pt\lower3pt\hbox{\epsffile{r3-#1.eps}}}}

%\def\i{\hbox{\it i}}

%%%%%%%%%%%%%%%%%%%%%%%%%%%%%%%%%%%%%%%%%%%%%%%%%%%%%%%%%%%%%%%%%%%%%%%%%%%%%

\def\l{\lambda}

\def\RR{{\Bbb R}}

\def\lra{\longrightarrow}
\def\ra{\rightarrow}

\def\arrowunder#1{\raise4pt\vtop{\baselineskip=0pt\lineskip=0pt
      \ialign{\hfill##\hfill\cr${\ss #1}$\cr$\lra$\cr}}}

\def\s{\sigma}

\def\xadj{\hbox{\sixbf adj}}

\def\xR{\hbox{\sixbf R}}

\def\leftbr{[\hskip-1.5pt[}
\def\rightbr{]\hskip-1.5pt]}

\def\dslash{\partial\hskip-5pt/}

\def\<{{<}}
\def\>{{>}}

\def\TableItem#1{%                             % centers item in ruled table
   \hfil\tablespace                             % left glue
   #1\relax                                     % item
   \tablespace                             % right glue
    }%
\thicksize=\thinsize

\def\dal{\hbox{\rlap{$\sqcap$}$\sqcup$}}

\def\Q{Q}

\def\lra{\longrightarrow}

\def\ra{\rightarrow}
\def\la{\leftarrow}

\def\rarrowover#1{\vtop{\baselineskip=0pt\lineskip=0pt
      \ialign{\hfill##\hfill\cr$\ra$\cr$#1$\cr}}}

\def\larrowover#1{\vtop{\baselineskip=0pt\lineskip=0pt
      \ialign{\hfill##\hfill\cr$\la$\cr$#1$\cr}}}

\def\<{{<}}
\def\>{{>}}

\def\xadj{\hbox{\sixbf adj}}

\def\xR{\hbox{\sixbf R}}

%%%%%%%%%%%%%%%%%%%%%%%%%%%%%%%%%%%%%%%%%%%%%%%%%%%%%%%%%%%%%%%%%%%%%%%%%%%%
%
%      References
%
%%%%%%%%%%%%%%%%%%%%%%%%%%%%%%%%%%%%%%%%%%%%%%%%%%%%%%%%%%%%%%%%%%%%%%%%%%%%

\ref\BaggerLambertI{J. Bagger and N. Lambert, {\xit ``Modeling
multiple M2's''}, \PRD{75}{2007}{045020} [\hepth{0611108}].}

\ref\BaggerLambertII{J. Bagger and N. Lambert, {\xit ``Gauge symmetry
and supersymmetry of multiple M2-branes''}, \PRD{77}{2008}{065008}
[\arxiv{0711}{0955}].} 

\ref\BaggerLambertIII{J. Bagger and N. Lambert, {\xit ``Comments on
multiple M2-branes''}, \JHEP{08}{02}{2008}{105} [\arxiv{0712}{3738}].}

\ref\Gustavsson{A. Gustavsson, {\xit ``Algebraic structures on
parallel M2-branes''}, \NPB{811}{2009}{66} [\arxiv{0709}{1260}].}

\ref\Papadopoulos{G. Papadopoulos, {\xit ``M2-branes, 3-Lie algebras
and Pl\"ucker relations''}, \jhep{08}{05}{2008}{054}
[\arxiv{0804}{2662}].}

\ref\GauntlettGutowski{J.P. Gauntlett and J.B. Gutowski, {\xit
``Constraining maximally supersymmetric membrane actions''},
\hfill\break\arxiv{0804}{3078}.} 

\ref\LambertTong{N. Lambert and D. Tong, {\xit ``Membranes on an
orbifold''}, \PRL{101}{2008}{041602} \hfill\break[\arxiv{0804}{1114}].}

\ref\DMPvR{J. Distler, S. Mukhi, C. Papageorgakis and M. van
Raamsdonk, {\xit ``M2-branes on M-folds''}, \jhep{08}{05}{2008}{038}
[\arxiv{0804}{1256}].}

\ref\StringTalks{Talks by N. Lambert, J. Maldacena and S. Mukhi at
Strings 2008, CERN, Gen\`eve, August 2008,
\hhref{www.cern.ch/strings2008};
talks by D. Jafferis and I. Klebanov at Strings 2009, Roma, June 2009,
\hhref{strings2009.roma2.infn.it}.}

\ref\CederwallNilssonTsimpisI{M. Cederwall, B.E.W. Nilsson and D. Tsimpis,
{\xit ``The structure of maximally supersymmetric super-Yang--Mills
theory---constraining higher order corrections''},
\jhep{01}{06}{2001}{034} 
[\hepth{0102009}].}

\ref\CederwallNilssonTsimpisII{M. Cederwall, B.E.W. Nilsson and D. Tsimpis,
{\xit ``D=10 super-Yang--Mills at $\ss O(\a'^2)$''},
\JHEP{01}{07}{2001}{042} [\hepth{0104236}].}

\ref\BerkovitsParticle{N. Berkovits, {\xit ``Covariant quantization of
the superparticle using pure spinors''}, \jhep{01}{09}{2001}{016}
[\hepth{0105050}].}

\ref\SpinorialCohomology{M. Cederwall, B.E.W. Nilsson and D. Tsimpis,
{\xit ``Spinorial cohomology and maximally supersymmetric theories''},
\jhep{02}{02}{2002}{009} [\hepth{0110069}];
M. Cederwall, {\xit ``Superspace methods in string theory, supergravity and gauge theory''}, Lectures at the XXXVII Winter School in Theoretical Physics ``New Developments in Fundamental Interactions Theories'',  Karpacz, Poland,  Feb. 6-15, 2001, \hepth{0105176}.}

\ref\Movshev{M. Movshev and A. Schwarz, {\xit ``On maximally
supersymmetric Yang--Mills theories''}, \NPB{681}{2004}{324}
[\hepth{0311132}].}

\ref\BerkovitsI{N. Berkovits,
{\xit ``Super-Poincar\'e covariant quantization of the superstring''},
\jhep{00}{04}{2000}{018} [\hepth{0001035}].}

\ref\BerkovitsNonMinimal{N. Berkovits,
{\xit ``Pure spinor formalism as an N=2 topological string''},
\jhep{05}{10}{2005}{089} [\hepth{0509120}].}

\ref\CederwallNilssonSix{M. Cederwall and B.E.W. Nilsson, {\xit ``Pure
spinors and D=6 super-Yang--Mills''}, \arxiv{0801}{1428}.}

\ref\CGNN{M. Cederwall, U. Gran, M. Nielsen and B.E.W. Nilsson,
{\xit ``Manifestly supersymmetric M-theory''},
\JHEP{00}{10}{2000}{041} [\hepth{0007035}];
{\xit ``Generalised 11-dimensional supergravity''}, \hepth{0010042}.
}

\ref\CGNT{M. Cederwall, U. Gran, B.E.W. Nilsson and D. Tsimpis,
{\xit ``Supersymmetric corrections to eleven-dimen\-sional supergravity''},
\jhep{05}{05}{2005}{052} [\hepth{0409107}].}

\ref\HoweTsimpis{P.S. Howe and D. Tsimpis, {\xit ``On higher order
corrections in M theory''}, \jhep{03}{09}{2003}{038} [\hepth{0305129}].}

\ref\NilssonPure{B.E.W.~Nilsson,
{\xit ``Pure spinors as auxiliary fields in the ten-dimensional
supersymmetric Yang--Mills theory''},
\CQG3{1986}{{\xrm L}41}.}

\ref\HowePureI{P.S. Howe, {\xit ``Pure spinor lines in superspace and
ten-dimensional supersymmetric theories''}, \PLB{258}{1991}{141}.}

\ref\HowePureII{P.S. Howe, {\xit ``Pure spinors, function superspaces
and supergravity theories in ten and eleven dimensions''},
\PLB{273}{1991}{90}.} 

\ref\FreGrassi{P. Fr\'e and P.A. Grassi, {\xit ``Pure spinor formalism
for OSp(N$\ss |$4) backgrounds''}, \arxiv{0807}{0044}.}

\ref\CederwallBLG{M. Cederwall, {\xit ``N=8 superfield formulation of
the Bagger--Lambert--Gustavsson model''}, \jhep{08}{09}{2008}{116} 
[\arxiv{0808}{3242}].}

\ref\CederwallABJM{M. Cederwall, {\xit ``Superfield actions for N=8 and N=6 conformal theories in three dimensions''}, \jhep{08}{10}{2008}{070} 
[\arxiv{0809}{0318}].}

\ref\BandosBLG{I. Bandos, {\xit ``NB BLG model in N=8 superfields''},
\PLB{669}{2008}{193} [\arxiv{0808}{3562}].} 

\ref\GranNilssonPetersson{U. Gran, B.E.W. Nilsson and C. Petersson,
{\xit ``On relating multiple M2 and D2-branes''}, 
\jhep{08}{10}{2008}{067} [\arxiv{0804}{1784}].}

\ref\MarneliusOgren{R. Marnelius and M. \"Ogren, {\xit ``Symmetric
inner products for physical states in BRST quantization''},
\NPB{351}{1991}{474}.} 

\ref\NilssonPalmkvist{B.E.W. Nilsson and J. Palmkvist, {\xit
``Superconformal M2-branes and generalized Jordan triple systems''},
\CQG{26}{2009}{075007} [\arxiv{0807}{5134}].} 

\ref\BaggerLambertIV{J. Bagger and N. Lambert, {\xit ``Three-algebras
and N=6 Chern--Simons gauge theories''}, 
\PRD{79}{2009}{025002} [\arxiv{0807}{0163}].}

\ref\BedfordBerman{J. Bedford and D. Berman, {\xit ``A note on quantum
aspects of multiple membranes''}, 
\PLB{668}{2008}{67} [\arxiv{0806}{4900}].}

\ref\GustavssonII{A. Gustavsson, {\xit ``One-loop corrections to
Bagger--Lambert theory''}, \NPB{807}{2009}{315} 
\hfill\break[\arxiv{0805}{4443}].}

\ref\BerkovitsICTP{N. Berkovits, {\xit ``ICTP lectures on covariant
quantization of the superstring''}, proceedings of the ICTP Spring
School on Superstrings and Related Matters, Trieste, Italy, 2002
[\hepth{0209059}.]} 

\ref\BandosTownsend{I. Bandos and P.K. Townsend, {\xit ``Light-cone M5
and multiple M2-branes''}, \CQG{25}{2008}{245003}
[\arxiv{0806}{4777}]; 
{\xit ``SDiff gauge theory and the M2 condensate''}, 
\jhep{09}{02}{2009}{013} [\arxiv{0808}{1583}].}

\ref\HoMatsuo{P.-M. Ho and Y. Matsuo, {\xit ``M5 from M2''},
\jhep{08}{06}{2008}{105} [\arxiv{0804}{3629}].}

\ref\SchnablTachikawa{M. Schnabl and Y. Tachikawa, {\xit ``Classification of
superconformal theories of ABJM type''}, \arxiv{0807}{1102}.}

\ref\ABJM{O. Aharony, O. Bergman, D.L. Jafferis and J. Maldacena,
{\xit ``N=6 superconformal Chern--Simons-matter theories, M2-branes
and their gravity duals''}, 
\jhep{08}{10}{2008}{091} [\arxiv{0806}{1218}].}

\ref\GomisMilanesiRusso{J. Gomis, G. Milanesi and J.G. Russo, {\xit
``Bagger--Lambert theory for general Lie algebras''}, 
\jhep{08}{06}{2008}{075} [\arxiv{0805}{1012}].}

\ref\BRTV{S. Benvenuti, D. Rodriguez-Gomez, E. Tonni and H. Verlinde,
{\xit ``N=8 superconformal gauge theories and M2 branes''}, 
\jhep{09}{01}{2009}{071} [\arxiv{0805}{1087}].}

\ref\HoImamuraMatsuo{P.-M. Ho, Y. Imamura and Y. Matsuo, {\xit ``M2 to
D2 revisited''}, \jhep{08}{07}{2008}{003} [\arxiv{0805}{1202}].}

\ref\BKKS{M. Benna, I. Klebanov, T. Klose and M. Smedb\"ack, {\xit
``Superconformal Chern--Simons theories and 
AdS${}_{\sss 4}$/ CFT${}_{\sss 3}$
correspondence''}, \jhep{08}{09}{2008}{072} [\arxiv{0806}{1519}].}

\ref\NishiokaTakayanagi{T. Nishioka and T. Takayanagi, {\xit ``On type
IIA Penrose limit and N=6 Chern--Simons theories''}, 
\jhep{08}{08}{2008}{001} [\arxiv{0806}{3391}].}

\ref\MinahanZarembo{J. Minahan and K. Zarembo, {\xit ``The Bethe
Ansatz for superconformal Chern-Simons''}, 
\jhep{08}{09}{2008}{040} [\arxiv{0806}{3951}].}

\ref\GaiottoWitten{D. Gaiotto and E. Witten, {\xit ``Janus
configurations, Chern--Simons couplings, and the theta-angle in N=4
super-Yang--Mills theory''}, \arxiv{0804}{2907}.}

\ref\HLLLPI{K. Hosomichi, K.-M. Lee, S. Lee, S. Lee and J. Park, {\xit
``N=4 superconformal Chern--Simons theories with hyper and twisted
hyper multiplets''}, \jhep{08}{07}{2008}{091} [\arxiv{0805}{3662}].}

\ref\HLLLPII{K. Hosomichi, K.-M. Lee, S. Lee, S. Lee and J. Park, {\xit
``N=5,6 superconformal Chern--Simons theories and M2-branes on
orbifolds''}, 
\jhep{08}{09}{2008}{002} [\arxiv{0806}{4977}].} 

\ref\MauriPetkou{A. Mauri and A.C. Petkou, {\xit ``An N=1 superfield
action for M2 branes''}, \PLB{666}{2008}{527} [\arxiv{0806}{2270}].}

\ref\CherkisSamann{S. Cherkis and C. S\"amann, {\xit ``Multiple
M2-branes and generalized 3-Lie algebras''}, 
\PRD{78}{2008}{066019} [\arxiv{0807}{0808}].}

\ref\BerkovitsNekrasovCharacter{N. Berkovits and N. Nekrasov, {\xit
``The character of pure spinors''}, \LMP{74}{2005}{75} [\hepth{0503075}].}
 
\ref\BerkovitsNekrasovReg{N. Berkovits and N. Nekrasov, {\xit
``Multiloop superstring amplitudes from non-minimal pure spinor
formalism''}, 
\jhep{06}{12}{2006}{029} [\hepth{0609012}].}

\ref\GrassiVanhove{P.A. Grassi and P. Vanhove, {\xit ``Higher-loop
amplitudes in the non-minimal pure spinor formalism''},
\jhep{09}{05}{2009}{089} [\arxiv{0903}{3903}].}

%%%%%%%%%%%%%%%%%%%%%%%%%%%%%%%%%%%%%%%%%%%%%%%%%%%%%%%%%%%%%%%%%%%%%%%%%%%%

\headtext={M. Cederwall: ``Pure spinor superfields...''}

\null
\vskip-1cm
\line{
\epsfxsize=18mm
\epsffile{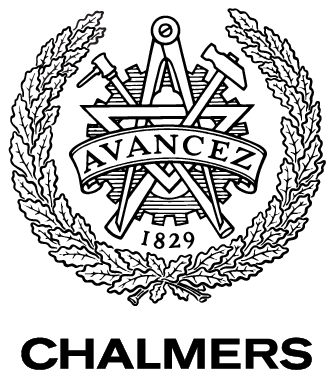}%Bildmarke30mm.eps}
\hfill}
\vskip-12mm
\line{\hfill G\"oteborg preprint}
%\line{\hfill hep-th/yymmnnn}
\line{\hfill June, {\old2009}}
\line{\hrulefill}

\vskip6\parskip

\centerline{\sixteenhelvbold
Pure spinor superfields,} 

\vskip3\parskip

\centerline{\sixteenhelvbold
with application to D=3 conformal models} 

\vskip6\parskip

\centerline{\twelvehelvbold
Martin Cederwall}

\vskip6\parskip

\centerline{\it Fundamental Physics}
\centerline{\it Chalmers University of Technology}
\centerline{\it SE 412 96 G\"oteborg, Sweden}
 \catcode`\@=11
\centerline{martin.cederwall@chalmers.se}
\catcode`\@=\active

\vskip6\parskip

{\narrower\noindent \underbar{Abstract:} 
I review and discuss the construction of supersymmetry multiplets and manifestly
supersymmetric Batalin--Vilkovisky actions using pure spinors, with emphasis on
models with maximal supersymmetry. The special cases of $D=3$, $N=8$
(Bagger--Lambert--Gustavsson) 
and $N=6$ (Aharony--Bergman--Jafferis--Maldacena) conformal models are
treated in detail. Most of the 
material is covered by the papers arXiv:0808.3242 and
arXiv:0809.0318. 
This is the written version of a talk given at 4th Baltic-Nordic
workshop ``Algebra, Geometry and Mathematical Physics'', Tartu,
Estonia, October 9-11, 2008, to appear in the Proceedings of the
Estonian Academy of Sciences, vol 4, 2010.

\noindent\underbar{Keywords:}
extended supersymmetry, superfields, pure spinors, conformal symmetry
\smallskip}
\vskip6\parskip

%\vtop{\baselineskip=.6\baselineskip\xxtt
%\line{\hrulefill}
%\catcode`\@=11
%\line{email: martin.cederwall@chalmers.se\hfill}
%\catcode`\@=\active
%}

%\eject

\noindent There is a close relation between supermultiplets and pure
spinors. The algebra of covariant fermionic derivatives in flat
superspace is generically of the form
$$
\{D_\a,D_\b\}=-T_{\a\b}{}^cD_c=-2\g_{\a\b}^cD_c\punkt\Eqn\Torsion
$$
If a bosonic spinor $\l^\a$ is {\it pure}, \ie, if the vector part
$(\l\g^a\l)$ of the spinor bilinear vanishes, the operator
$\Q=\l^\a D_\a$ becomes nilpotent, and may be used as a BRST operator.
This is, schematically, the starting point for pure spinor
superfields. (The details of course depend on the actual space-time and
the amount of supersymmetry. The pure spinor constraint may need to be
further specified. Eq. (\Torsion) may also contain more terms, due to
super-torsion and curvature.) The cohomology of $\Q$ will consist of
supermultiplets, which in case of maximal supersymmetry are
on-shell. The idea of manifesting maximal supersymmetry off-shell by using pure
spinor superfields $\Psi(x,\theta,\l)$ is to find an action whose
equations of motion is $\Q\Psi=0$.

The fact that pure spinors had a r\^ole to play in maximally
supersymmetric models was recognised early by Nilsson [\NilssonPure]
and Howe [\HowePureI,\HowePureII].
Pure spinor superfields were developed with the purpose of covariant
quantisation of superstrings by Berkovits
[\BerkovitsI,\BerkovitsParticle,\BerkovitsICTP,\BerkovitsNonMinimal] and the
cohomological structure was independently discovered in supersymmetric
field theory and supergravity, originally in the context of
higher-derivative deformations
[\CederwallNilssonTsimpisI,\CederwallNilssonTsimpisII,\SpinorialCohomology,\Movshev,\CederwallNilssonSix,\CGNN,\CGNT,\HoweTsimpis].
The present lecture only deals with pure spinors for maximally
supersymmetric field theory.

The canonical example of pure spinors is in $D=10$. There is only one
non-gravitational supermultiplet, namely super-Yang--Mills, so
this is what we expect to obtain. Expanding a
field $\Psi(x,\theta,\lambda)$ in powers of $\l$, one has
$$
\Psi(x,\theta,\lambda)=\sum_{n=0}^\infty\l^{\a_1}\ldots\l^{\a_n}
\psi_{\a_1\ldots\a_n}(x,\theta)
\punkt\Eqn\LambdaExpansion
$$ 
The implementation of the pure spinor constraint is as an abelian
gauge symmetry, where the generators $(\l\g^a\l)$ act
multiplicatively. The field $\Psi$ is defined modulo the ideal
generated by the constraint. A ``canonical'' representative of the
gauge orbits is provided by superfields
$\psi_{\a_1\ldots\a_n}(x,\theta)$ which, in addition to being
symmetric, are completely $\g$-traceless, \ie, in the modules
$(000n0)$ of the Lorentz algebra (where $\l^\a$ is in $(00001)$ and
$D_\a$ in $(00010)$, the two spinor chiralities).

In order to calculate the cohomology, we start by finding the
cohomology of zero-modes, $x$-independent fields. This cohomology is
easy to calculate (a purely algebraic calculation), and gives
information about the full cohomology.
It is worth noting that the zero-mode cohomology (which clearly would
have been empty for an unconstrained $\l$) may be read off from the
partition function for a pure spinor. It 
is in one-to-one
correspondence (for a concrete explanation of this fact, using the
reducibility of the pure spinor constraint,
see the appendix of ref. [\BerkovitsParticle] and
ref. [\BerkovitsNekrasovCharacter]) with
the six terms in the nominator of the partition function
$$
Z(t)={1-10t^2+16t^3-16t^5+10t^6-t^8\over(1-t)^{16}}
={(1+t^2)(1+4t+t^2)\over(1-t)^{11}}\punkt\Eqn\TenPartition
$$
(This partition function only counts the dimension of the space of
monomials is $\l$ with degree of homogeneity $p$ as the coefficient of
$t^p$. A more refined partitions function, specifying the actual
Lorentz modules appearing, can of course be written down; for this I refer to
ref. [\BerkovitsNekrasovCharacter].)
The zero-mode cohomology is illustrated in Table 1. There, each column
represent a field in the expansion (\LambdaExpansion), and the
vertical direction is the expansion in $\theta$. The columns have been
shifted so that components on the same row have the same dimension,
\ie, so that $Q$ acts horizontally. Since $\l$ carries ghost number 1
and dimension $-1/2$,
the component field $\psi_{\a_1\ldots\a_n}$ has ghost number
$\hbox{gh}(\Psi)-n$ and dimension $\hbox{dim}(\Psi)+{n\over2}$. 
It is natural to let $\hbox{gh}(\Psi)=1$ and $\hbox{dim}(\Psi)=0$
and take $\Psi$ to be fermionic. Then
the scalar (ghost number 1, dimension 0) in the first column is
interpreted as the Yang--Mills 
ghost and the vector and spinor in the second column as the fields of the
super-Yang--Mills multiplet (the field $\psi_\a$ of ghost number 0 and
dimension $1/2$ is the lowest-dimensional connection component $A_\a$
in a superfield treatment of super-Yang--Mills). The remaining fields
are the corresponding 
antifields, in the Batalin--Vilkovisky (BV) sense. It is striking that one
inevitably is lead to the BV formalism. It of course exists also in a
component formalism, but when one uses pure spinors it is not
optional. This means that any action formed in this formalism will be
a BV action, and that the appropriate consistency relation (encoding the
generalised gauge symmetry) is the master equation.

$$
\vtop{\baselineskip20pt\lineskip0pt
\ialign{
$\hfill#\quad$&$\,\hfill#\hfill\,$&$\,\hfill#\hfill\,$&$\,\hfill#\hfill\,$
&$\,\hfill#\hfill\,$&$\,\hfill#\hfill$&\qquad\qquad#\cr
            &n=0    &n=1    &n=2    &n=3    &n=4\cr
\hbox{dim}=0&(00000)&       &       &       &\phantom{(00000)}       \cr
        \fr2&\bullet&\bullet&               &       \cr 
           1&\bullet&(10000)&\bullet&       &       \cr
       \Fr32&\bullet&(00001)&\bullet&\bullet&       \cr
           2&\bullet&\bullet&\bullet&\bullet&\bullet\cr
       \Fr52&\bullet&\bullet&(00010)&\bullet&\bullet\cr
           3&\bullet&\bullet&(10000)&\bullet&\bullet\cr
       \Fr72&\bullet&\bullet&\bullet&\bullet&\bullet\cr
           4&\bullet&\bullet&\bullet&(00000)&\bullet\cr
       \Fr92&\bullet&\bullet&\bullet&\bullet&\bullet\cr
}}
$$

%\vskip2\parskip
\centerline{\it Table 1. The cohomology of the $D=10$ SYM complex.}

\vskip2\parskip
To go from the zero-mode cohomology to the complete cohomology, one
easily convinces oneself that component fields in the modules
contained in the zero-mode cohomology will be subject to differential
constraints in the modules of the zero-mode cohomology in the next
column to the right. This gives the proper relations for the
linearised on-shell super-Yang--Mills multiplet. (If a multiplet is
an off-shell representation of supersymmetry, as is generically the
case for half-maximal or lower supersymmetry, there will consequently be no
anti-fields in the cohomology. These instead come in a separate pure
spinor superfield [\CederwallNilssonSix].)

This far, we have not considered the actual solutions of pure spinor
constraint, but rather regarded the pure spinor as a book-keeping
device. When one wants to write down an action, this is no longer
possible. For an action, a measure is needed. The linearised action should be
``$\int\Psi Q\Psi$" for some suitable definition of
``$\int$''. Clearly, ``$\int$'' must have ghost number $-3$. In the
cohomology, there is a singlet at $\l^3\theta^5$. Defining a measure
as a ``residue'', picking the corresponding component, has the right
ghost number, and also the correct dimension. However, it is
singular, so components of $\Psi$ with high enough power in $\l$ or
$\theta$ drop out of the putative action defined in this manner, and
the equation of motion $Q\Psi=0$ does not follow. Still, the
corresponding tensorial structure can be used for an invariant integral
over $\lambda$. It is clear from the partition function
(\TenPartition) that $\lambda$ contains 11 degrees of freedom (out of
the 16 for an unconstrained spinor). Explicit solution of the pure
spinor constraint also shows that when imposed on a {\it complex} spinor,
only 5 out of the ten constraints are independent (see \eg\
[\BerkovitsParticle] for details). Defining the scalar at
$\l^3\theta^5$ as
$T_{\a_1\a_2\a_3}^{\b_1\ldots\b_{11}}\e_{\b_1\ldots\b_{16}}\l^{\a_1}\l^{\a_2}\l^{\a_3}\theta^{\b_{12}}\theta^{\b_{13}}\theta^{\b_{14}}\theta^{\b_{15}}\theta^{\b_{16}}$,
where $T$ thus is a Lorentz invariant tensor, one defines the conjugate
invariant tensor $\tilde T^{\a_1\a_2\a_3}{}_{\b_1\ldots\b_{11}}$, and
the integration is
$$
[d\l]\l^{\a_1}\l^{\a_2}\l^{\a_3}\sim\tilde
T^{\a_1\a_2\a_3}{}_{\b_1\ldots\b_{11}}d\l^{\a_1}\wedge\ldots\wedge d\l^{\a_{11}}
\punkt\Eqn\TenLambdaInt
$$
In ref. [\BerkovitsNonMinimal], Berkovits solved the problem how to
make sense of this integration and using it as part of a non-singular
measure for the pure spinor superspace. The solution involves a
non-minimal set of pure spinor variables, which in addition to $\l^\a$
contains a bosonic conjugate spinor $\bar\l_\a$ (which in euclidean signature
can be viewed as the complex conjugate of $\l^\a$) obeying
$(\bar\l\g^a\bar\l)=0$ and a fermionic spinor $r_\a$ with
$(\bar\l\g^ar)=0$. The new BRST operator is 
$\Q=\l^\a D_\a+{\*\over\*\bar\l_\a}r_\a$, and its
cohomology is independent of $\bar\l$ and $r$. One assigns ghost
number $-1$ and dimension $1/2$ to $\bar\l$ and ghost number $0$ and
dimension $1/2$ to $r$. The measure for $\bar\l$ is
the complex conjugate to the one defined in eq. (\TenLambdaInt) for
$\l$, and for $r$:
$$
[dr]\sim{\star}\tilde T^{\a_1\a_2\a_3}{}_{\b_1\ldots\b_{11}}
\bar\l_{\a_1}\bar\l_{\a_2}\bar\l_{\a_3}
{\*\over\*r_{\b_1}}\ldots{\*\over\*r_{\b_{11}}}
\punkt\eqn
$$
Using these integration measures, and the ordinary ones for $x$ and
$\theta$, we list the dimensions and ghost numbers for the theory
after dimensional reduction to $D$ dimensions in Table 2.
So, the ghost numbers match, and also the dimensions (${1\over g^2}$
has dimension $D-4$ in $D$ dimensions). 
%irrespectively of the
%assignments of $\hbox{dim}(\m)$ and $\hbox{gh\#}(\mu)$.
 
\vskip3\parskip
\thicksize=\thinsize
\ruledtable
                |gh\#           |dim            \cr
$d^Dx$          |$0$            |$-D$             \crnorule
$d^{16}\theta$  |$0$            |$8$              \crnorule
$[d\l]$         |$8$            |$-4$             \crnorule
$[d\bar\lambda]$         |$-8$   |$4$    \crnorule
$[dr]$          |$-3$  |$-4$      \cr
total           |$-3$           |$-(D-4)$         
\endruledtable
\vskip2\parskip
\centerline{\it Table 2. The dimensions and ghost numbers of the
$D=10$ measure.}
\vskip2\parskip

The $\l$ and $\bar\l$ integrations are non-compact and need
regularisation. In ref. [\BerkovitsNonMinimal] this is achieved,
following ref. [\MarneliusOgren], by the insertion of a factor 
$N=e^{\{\Q,\chi\}}$. Since this differs from 1 by a $\Q$-exact term,
the regularisation is independent of the choice of the 
fermion $\chi$. The choice $\chi=-\bar\l_\a\theta^\a$ gives 
$N=e^{-\l^\a\bar\l_\a-r_\a\theta^\a}$ and regularises the bosonic
integrations at infinity. At the same time, it explains how the term
at $\theta^5$ is picked out, this follows after integration over $r$.
$N$ has definite ghost number 0 for the assignments for ghost number
and dimension above (although
any other assignment gives the correct ghost number and dimension for
the non-$Q$-exact part).

An action for ten-dimensional super-Yang--Mills (or any dimensional
reduction) can now be written in the Chern--Simons-like form 
[\BerkovitsParticle]
$$
S={1\over 2g^2}\int\<\Psi,Q\Psi+\fr3[\Psi,\Psi]\>_{\xadj}
\punkt\Eqn\TenLagrangian
$$
Note that there is no 4-point coupling. The component field 4-point
coupling arises after elimination of unphysical components.
One must however remember that this is a classical BV action. It obeys
the classical master equation $(S,S)=0$, where the anti-bracket takes
the simple form
$$
(A,B)=\int A\<{\larrowover\delta\over\delta\Psi(Z)}[dZ]
{\rarrowover\delta\over\delta\Psi(Z)}\>_{\xadj} B\punkt\eqn
$$
In order to perform quantum calculations with path integral over
$\Psi$, gauge fixing has to be implemented. This involves traditional
gauge fixing (of the component gauge field) as well as elimination of
the anti-fields. I will comment briefly on gauge fixing towards the
end of the lecture.

As already mentioned, pure spinor formulations are relevant for BV
action formulations of any maximally supersymmetric model (exceptions
being models containing self-dual tensors). I would now like to
illustrate how they may be used for 3-dimensional conformal
models. The pure spinor actions turn out to have a much simpler
structure than the component actions.
There has recently been much interest in conformal
three-dimensional theories. Following the discovery of the existence
of a maximally supersymmetric ($N=8$) interacting theory of scalar
multiplets coupled to Chern--Simons, the Bagger--Lambert--Gustavsson
(BLG) theory [\BaggerLambertI,\Gustavsson,\BaggerLambertII,\BaggerLambertIII], 
much effort has been spent on trying to generalise the
construction and to interpret it in terms of an AdS boundary model of
multiple M2-branes. The interesting, but restrictive, algebraic
structure of the model, containing a 3-algebra with antisymmetric
structure constants, turned out to have only one finite-dimensional
realisation [\Papadopoulos,\GauntlettGutowski], 
possible to interpret in term of two M2-branes [\LambertTong,\DMPvR] (see
however refs. [\HoMatsuo,\BandosTownsend] 
dealing with the infinite-dimensional solution
related to volume-preserving diffeomorphisms in three dimensions).

It then became an urgent question how the stringent requirements in
the BLG theory could be relaxed. There are different
possibilities. One may let the scalar product on the matter
representation be degenerate [\GranNilssonPetersson]. 
This works at the level of equations
of motion, but does not allow for an action principle. One may also go
one step further, and add further null directions to that degenerate
case, which leads to scalar products with indefinite signature 
[\GomisMilanesiRusso,\BRTV,\HoImamuraMatsuo] (and
consequently to matter kinetic terms with different signs). Or, finally,
one may reduce the number of supersymmetries, specifically to $N=6$,
as proposed by Aharony, Bergman, Jafferis and Maldacena (ABJM)
[\ABJM], or maybe even to lower $N$ [\GaiottoWitten,\HLLLPI].
The $N=6$ models were further studied in refs. 
[\BKKS,\NishiokaTakayanagi,\MinahanZarembo,\HLLLPII,\BaggerLambertIV,\SchnablTachikawa,\NilssonPalmkvist]
(among other papers).  
For recent developments in the theory of multiple membranes, we refer
to ref. [\StringTalks] and references given there. The literature on
the subject is huge, and we apologise for omissions of references
to relevant papers.

%{\it Here comes a presentation of the pure spinors in D=3, the
%constraint structure, partition functions, cohomologies etc. Partly
%from below.}

%{\it Comment on the interesting feature that $D=3$ CS has $N=8$
%supersymmetry (trivial on-shell).}

The superfield
formulation of the BLG model was given in ref.
[\CederwallBLG] (see also 
ref. [\BandosBLG], where the on-shell superfields were constructed for the
example of the BLG model based on the
infinite-dimensional algebra of volume-preserving diffeomorphisms in
three dimensions). A superfield formulation with $N=1$ superfields was given in
ref. [\MauriPetkou] and with $N=2$ superfields in ref. 
[\CherkisSamann]. In ref. [\CederwallBLG]  
we constructed an action in an $N=8$ pure spinor superspace
formulation of the BLG model, which covers all situations with $N=8$
above except the ones with degenerate scalar product. The construction
was essentially performed using minimal pure spinor variables, and the
issue of the integration measure was more or less neglected (the
measure was assumed to exist).
In the subsequent paper [\CederwallABJM] also the $N=6$ ABJM models
were treated, and integration measures were defined using non-minimal
variables for both types of models.
 
Let us first briefly review the results of ref. [\CederwallBLG]. Since the BLG
model is maximally supersymmetric, component formulations and also
usual superspace formulations are on-shell. There is no finite set of
auxiliary fields. A pure spinor treatment is necessary in order to
write an action in a generalised BRST setting. 

The Lorentz algebra in $D=3$ is $so(1,2)\approx sl(2,\RR)$. The $N=8$
theory has an $so(8)$ R-symmetry, and we choose the fermionic
coordinates and derivatives to transform as $({\bf 2},{\bf
8}_s)=(1)(0010)$ under $sl(2)\oplus so(8)$. This representation is
real and self-conjugate. The pure spinors transform in the same
representation, and are written as $\l^{A\a}$, where $A$ is the
$sl(2)$ index and $\a$ the $so(8)$ spinor index. As usual, a BRST
operator is formed as $Q=\l^{A\a}D_{A\a}$, $D$ being the fermionic
covariant derivative. The nilpotency of $Q$ demands that
$$
(\l^A\l^B)=0\komma\Eqn\EightConstraint
$$
where $(\ldots)$ denotes contraction of $so(8)$ spinor indices, 
since the superspace torsion has to be projected
out. This turns out to be the full constraint\foot\star{The
vanishing of the ``torsion representation'' --- the vector part of the
spinor bilinear --- is necessary, but does not
always give the full pure spinor constraint. One example where further
constraints are needed is $N=4$, $D=4$ super-Yang--Mills theory.}. As
will soon be clear, it is essential that not only
$(\l^A\s_{IJKL}\l^B)$ but also $\e_{AB}(\l^A\s_{IJ}\l^B)$ is left non-zero. 
These pure spinors are similar to those encountered in
ref. [\FreGrassi]. 

The
``pure spinor wave function'' for the Chern--Simons field is a
fermionic scalar $\Psi$ of (mass) dimension 0 and ghost number 1. For the
matter multiplet we have a bosonic field $\Phi^I$ in the $so(8)$ vector
representation $(0)(1000)$ of dimension 1/2 and ghost number 0. In
addition to the pure spinor constraint, the matter field is identified
modulo transformations
$$
\Phi^I\rightarrow\Phi^I+(\l^A\sigma^I\varrho_A)\Eqn\PhiReducibility
$$ 
for arbitrary
$\varrho$. (This type of additional gauge invariance is typical for
fields in some non-trivial module of the structure group. Without it,
the cohomology would be the tensor product of the module with the
cohomology of a field in the trivial module.) 

In
this minimal pure spinor formulation the fields are expanded in power
series in $\lambda$, \ie, in decreasing ghost number. 
The pure spinor partition function is easily calculated to be 
$$
Z_1(t)={1-3t^2+3t^4-t^6\over(1-t)^{16}}={(1+t)^3\over(1-t)^{13}}
\punkt\Eqn\ThreePartitionOne
$$
The partition for a matter field is
$$
Z_8={8-16t+16t^3-8t^4\over(1-t)^{16}}=8{(1+t)\over(1-t)^{13}}
\punkt\Eqn\ThreePartitionEight
$$
These expressions seem to imply that the number of independent degrees
of freedom of a pure spinor is 13, \ie, that the pure spinor
constraint in this case is irreducible. This is verified by a concrete
solution of the constraint (for a complex $\l$) [\CederwallABJM]. 
As for the $D=10$
pure spinors, the partition functions can of course be refined to
include not only number of fields, but also modules of the structure group.

The field content (ghosts, fields and their antifields) is read off from
the zero-mode BRST cohomology given in Tables 3 and 4 for the
Chern--Simons and matter sectors respectively.
\vskip2\parskip
\vbox{
$$
\vtop{\baselineskip20pt\lineskip0pt
\ialign{
$\hfill#\quad$&$\,\hfill#\hfill\,$&$\,\hfill#\hfill\,$&$\,\hfill#\hfill\,$
&$\,\hfill#\hfill\,$&$\,\hfill#\hfill\,$\cr
\hfill\hbox{gh\#}=&1    &0    &-1    &-2  &-3 \cr
\hbox{dim}=0&(0)(0000)&\phantom{(0)(0000)}&       &          \cr
        \Fr12&\bullet&\bullet&\phantom{(0)(0000)}   \cr
           1&\bullet&(2)(0000)&\bullet&\phantom{(0)(0000)} \cr
       \Fr32&\bullet&\bullet&\bullet&\bullet&\phantom{(0)(0000)}     \cr
           2&\bullet&\bullet&(2)(0000)&\bullet&\bullet\cr
       \Fr52&\bullet&\bullet&\bullet&\bullet&\bullet\cr
           3&\bullet&\bullet&\bullet&(0)(0000)&\bullet\cr
       \Fr72&\bullet&\bullet&\bullet&\bullet&\bullet\cr
%           4&\bullet&\bullet&\bullet&\bullet&\bullet\cr
}}
$$
\vskip2\parskip
%\nobreak
\centerline{\it Table 3. The cohomology of the scalar complex.}
}

\vskip2\parskip
\vbox{
$$
\vtop{\baselineskip20pt\lineskip0pt
\ialign{
$\hfill#\quad$&$\,\hfill#\hfill\,$&$\,\hfill#\hfill\,$&$\,\hfill#\hfill\,$
&$\,\hfill#\hfill\,$&$\,\hfill#\hfill\,$\cr
\hfill\hbox{gh\#}=&0    &-1    &-2    &-3  &-4 \cr
\hbox{dim}=\Fr12&(0)(1000)&\phantom{(0)(0000)}&       &          \cr
        1&(1)(0001)&\bullet&\phantom{(0)(0000)}   \cr
           \Fr32&\bullet&\bullet&\bullet&\phantom{(0)(0000)} \cr
       2&\bullet&(1)(0001)&\bullet&\bullet&\phantom{(0)(0000)}     \cr
           \Fr52&\bullet&(0)(1000)&\bullet&\bullet&\bullet\cr
       3&\bullet&\bullet&\bullet&\bullet&\bullet\cr
           \Fr72&\bullet&\bullet&\bullet&\bullet&\bullet\cr
}}
$$
\vskip2\parskip
%\nobreak
\centerline{\it Table 4. The cohomology of the vector complex.}
}
\vskip2\parskip

\noindent We observe that the field content is the right one. In
$\Psi$ we find the ghost, the gauge connection, its antifield and the
antighost. The antifield has dimension 2 (as opposed to \eg\ $D=10$
super-Yang--Mills, where it has dimension 3), indicating equations of
motion that are first order in derivatives. 
It is quite striking that the (bosonic) Chern--Simons model has a
natural supersymmetric off-shell extension, although the supersymmetry
becomes trivial on-shell. It is not meaningful to talk about a gaugino
field. 
In $\Phi$ we find the
eight scalars $\phi^I$, the fermions $\chi^{A\dot\a}$ and their
antifields. In addition, the field $\Psi$ transforms in the adjoint
representation 
{\bf adj} of some gauge group and $\Phi^I$ in some representation {\bf
R} of the gauge group. The corresponding indices are suppressed.

In order to derive the equations of motion for the physical component fields,
one starts from the ghost number zero part of the fields (\ie,
$\Phi^I\rightarrow\phi^I(x,\theta)$ and $\Psi\rightarrow\l^\a A_\a(x,\theta)$
respectively), and examines the content of the $\theta$ expansion by
repeated application of fermionic covariant derivatives, using the
pure spinor constraint and the reducibility (\PhiReducibility) when
they occur. As a guideline one has the cohomology at ghost number one;
these representations are the only ones where an equation of motion
may sit, for obvious reasons. In this manner, one derives the
linearised component equations $\dal\phi^I=0$, $\dslash\chi^A=0$ for the
scalar multiplet, and $dA=0$ for the Chern--Simons field, and also the
interacting equations from the actions below. 

In ref. [\CederwallBLG], it was assumed that a non-degenerate measure
can be formed using a non-minimal extension of the pure spinor
variables along the lines of ref. [\BerkovitsNonMinimal]. This measure,
including the three-dimensional integration, should carry dimension 0
and ghost number $-3$, and should allow ``partial integration'' of the
BRST charge $Q$. It was then shown that the Lagrangian of the
interacting model is of a very simple form, containing essentially a
Chern--Simons like term for the Chern--Simons field, minimally coupled
to the matter sector:
$$
S=\int\<\Psi,Q\Psi+\fr3[\Psi,\Psi]\>_{\xadj}
    +\int\fr2M_{IJ}\<\Phi^I,Q\Phi^J+\Psi\cdot\Phi^J\>_{\xR}\punkt\Eqn\Lagrangian
$$
The brackets denote (non-degenerate) scalar products on {\bf adj} and
{\bf R}, $[\cdot,\cdot]$ the Lie bracket of the gauge algebra and
$T\cdot x$ the action of the Lie algebra element in the representation
{\bf R}. $M_{IJ}$ is the pure spinor bilinear
$\e_{AB}(\l^A\sigma_{IJ}\l^B)$, which is needed for several reasons: in order to
contract the indices on the $\Phi$'s antisymmetrically, to get a
Lagrangian of ghost number 3, and to ensure invariance in the
equivalence classes defined by eq. (\PhiReducibility).

The invariances of the interacting theory (equivalent to the classical
master equation $(S,S)=0$), generalising the BRST
invariance in the linearised case, are:
$$
\eqalign{
\d\Psi&=Q\Psi-[\Lambda,\Psi]-M_{IJ}\{\Phi^I,\Xi^J\}\komma\hfill\cr
\d\Phi^I&=-\Lambda\cdot\Phi^I+(Q+\Psi\cdot)\Xi^I\komma\hfill\cr}
\eqn
$$
where $\Lambda$ is an adjoint boson of dimension 0 and ghost number 0,
and $\Xi^I$ a fermionic vector in {\bf R} of dimension 1/2 and ghost number
$-1$. Here we also introduced the bracket $\{\cdot,\cdot\}$ for the
formation of an adjoint from the antisymmetric product of two elements
in {\bf R}, defined via 
$\<x,T\cdot y\>_{\xR}=\<T,\{x,y\}\>_{\xadj}$. 
The invariance with parameter $\Lambda$ is manifest. The
transformation with $\Xi$ has to be checked. One then finds that the
transformation of the matter field $\Phi$ gives a ``field strength''
contribution from the anticommutator of the two factors $Q+\Psi$,
which is cancelled against the variation of the Chern--Simons
term. The single remaining term comes from the transformation of the
$\Psi$ in the covariant matter kinetic term, and it is proportional to 
$M_{IJ}M_{KL}\<\{\Phi^I,\Phi^J\},\{\Phi^K,\Xi^L\}\>_{\xadj}$. Due to the
pure spinor constraint, $M_{[IJ}M_{KL]}=0$. This was shown in
ref. [\CederwallBLG], using the simple observation that the only
$sl(2)$ singlet at the fourth power of $\l$ is in the $sl(2)\oplus
so(8)$ representation $(0)(0200)$ --- the four-index antisymmetric
tensors $(0)(0020)$ or $(0)(0002)$ do not occur. So if the structure
constants of the 3-algebra defined by
$\<\{x,a\},\{b,c\}\>_{\xadj}=\<x,\leftbr a,b,c\rightbr\>_{\xR}$ are
antisymmetric, this term vanishes. It was also checked that the
commutator of two $\Xi$-transformations gives a
$\Lambda$-transformation together with a transformation of the type
(\PhiReducibility). In this way, one is naturally led to the 3-algebra
structure with a minimal amount of input, essentially a ``minimal
coupling''. I would like to stress that although the pure spinor
action contains at most 3-point couplings, the full component action
with up to 6-point interactions will arise when unphysical component fields are
eliminated. 

For the $N=6$ ABJM models, the results are very similar.
Due to lack of time and space, I will not go into details, but refer
to ref. [\CederwallABJM]. The end result is a weaker condition on the
structure constants of the ``3-algebra'', which is just the
appropriate one [\BaggerLambertIV]. 
The classification of such algebraic structures was
performed in ref. [\SchnablTachikawa].
It is satisfactory that the structure of the pure spinors in both
cases give the necessary and sufficient algebraic structure by the
vanishing of a single term in the transformation of minimally coupled matter.

In both the $N=8$ and $N=6$ theories in $D=3$, the na\"\i ve measure
sits at $\l^3\theta^3$. In analogy with the ten-dimensional case, we
need the number of irreducible constraints on the pure spinors to
equal the number of $\theta$'s. Indeed, the constraints, which in both
cases sit in the vector representation of $so(1,2)$, turn as mentioned
out to be
irreducible, which is straightforward to check (explicit solutions
were given in ref. [\CederwallABJM]). 

We can write the invariant tensors as 
$$
\eqalign{
&\e_{abc}(\l\g^a\theta)(\l\g^b\theta)(\l\g^c\theta)\cr
&\qquad=T_{(A_1\a_1,A_2\a_2,A_3\a_3)[B_1\b_1,B_2\b_2,B_3\b_3]}
\l^{A_1\a_1}\l^{A_2\a_2}\l^{A_3\a_3}
        \theta^{B_1\b_1}\theta^{B_2\b_3}\theta^{B_3\b_3}\cr}
\eqn
$$ 
in the $N=8$ case, and as a similar  expression when $N=6$.
The integration measures are constructed using these invariant tensors
in a manner completely analogous to the measure in $D=10$.

Let us examine the dimensions and ghost numbers of the total
measures. The analogies of Table 2 are obtained by simple counting
(the $N=6$ case is included 
for completeness):
\vskip3\parskip
\ruledtable
\dbl\multispan2 \hfil$N=8$ \hfil     \dbl\multispan2 \hfil$N=6$\hfil\cr
               \dbl gh\#   |dim  \dbl gh\#    |dim    \cr
$d^3x$         \dbl$0$            |$-3$             
               \dbl$0$             |$-3$               \crnorule
$[d\theta]$    \dbl$0$            |$8$              
               \dbl$0$             |$6$               \crnorule
$[d\l]$         \dbl$10$            |$-5$            
               \dbl$6$             |$-3$               \crnorule
$[d\bar\l]$         \dbl$-10$          |$5$    
               \dbl$-6$             |$3$               \crnorule
$[dr]$          \dbl$-3$            |$-5$      
               \dbl$-3$             |$-3$               \cr
total           \dbl$-3$           |$0$  
                \dbl$-3$           |$0$         
\endruledtable
\vskip2\parskip
\centerline{\it Table 5. The dimensions and ghost numbers of the
$N=8$ and $N=6$ measures.}
\vskip2\parskip
\noindent In both cases 
we get a non-degenerate measure of dimension 0 and ghost
number $-3$, as desired for a conformal theory. 
Also here, the measures of course have to be
regularised in the same way as in ref. [\BerkovitsNonMinimal]. We
insert a factor $N=e^{\{{\cal Q},\chi\}}$, where 
$\chi=-\m_{A\a}\theta^{A\a}$ for $N=8$ (and similarly for $N=6$).

To conclude, we have presented manifestly supersymmetric
formulations of the $N=8$ BLG models and the $N=6$ ABJM models. We have
also performed a detailed analysis of the pure spinor constraints and
provided proper actions based on non-degenerate measures on
non-minimal pure spinor spaces. We hope that these formulations may be
helpful in the future, \eg\ for the investigation of quantum
properties [\GustavssonII,\BedfordBerman] of the models.
In order to perform path integrals, one has to gauge fix. Gauge fixing
in the pure spinor formalism for the superparticle includes the
``$b$-ghost'', with the property $\{Q,b\}=\dal$. The $b$-ghost is a
composite operator in the pure spinor formalism, since $p^2=0$ is not
an independent constraint. This operator has singularities at $\l=0$,
which need to be regularised. Proposals for resolving this issue and
allowing for calculation of string amplitudes at arbitrary loop level 
have been made in
refs. [\BerkovitsNekrasovReg,\GrassiVanhove], but lead to complicated
expressions. It is possible that some simpler approach exists. 

I believe that much more is to be learnt from pure spinor superspace
formulations, especially of maximally supersymmetric theories. One very
interesting example, largely unexplored, is the issue of such
formulations of supergravities. I think that the treatment of the
scalar multiplet actions in the present work may contains clues to
supergravity, both considering the extra gauge invariances and the
extra factors of $\l$ in the action. Work is in progress.

%{\it Conclude, comment about gauge fixing etc. Also mention something
%about supergravity, and maybe U-duality.}

\acknowledgements The author would like to thank Bengt E.W. Nilsson,
Ulf Gran, Dimitrios Tsimpis, Pierre Vanhove, Nathan Berkovits, Niclas
Wyllard and Pietro Antonio Grassi for discussions and comments.

\refout

\end